# Regenerative Soot-II: Emission of carbon clusters from sooting plasma

**Shoaib Ahmad**

*National Centre for Physics, Quaid-i-Azam University Campus, Shahdara Valley, Islamabad, 44000, Pakistan*

Email: sahmad.ncp@gmail.com

**Abstract.** A hollow cathode source with a specially designed cusp magnetic field $B_z(r, \theta)$ provides clustering environment to the sputtered carbon atoms and ions. Source's distinctive features are its sooting properties that depend upon the two well-defined regions within the discharge. The cusp field drives both of these regions but in different modes – one region ionizes the source gas, introduces the sputtered C from the graphite hollow cathode while the other is the collision dominated gyration motion of the $C^+$ ions along $B_z(r, \theta)$ contours. Tunability of plasma's discharge characteristics provides the tool to investigate clustering in the sooting plasma

## 1 Introduction

Various techniques exist for the production of soot that lead to the formation of carbon clusters including chains, rings, fullerenes, onions and nanotubes [1–7]. Understanding of the mechanisms of clustering may lie in the synthesis of common features of the widely different physical methods of producing soot. The aim of the present study is the design of a tunable carbon plasma source. Such a source Nilorotron-I has been designed [8]. In this communication we describe the distinctive feature – its soot formation properties. Cold cathode uno- and duo-plasmatron sources [9–11] with hollow cathodes operating in Penning discharge mode with axial magnetic fields yield high current densities of singly and multiply charged gaseous ions. In the present design of the hollow cathode source we utilize the radial and azimuthal variations of hexapole magnetic field generating line cusps that enhance the phenomenon of the cathode wall sputtering by the support gas ions. Wall sputtering has been used as the means to introduce C into the plasma. A steady stream of carbon atoms is sputtered into the glow discharge plasma from graphite hollow cathode surface. The key to the ignition and sustenance of the discharge at low neon pressures $\sim 10^{-3}$ mbar is a set of six bar magnets wrapped around the hollow cathode providing an extended set of cusp magnetic field contours. Whereas, molecular gases $CO_2$, $CH_4$, $N_2$, noble gases i.e., Xe, Kr, Ar and Ne as well as mixtures Ar+$N_2$ and Ne+$N_2$ have been used, Ne has proved to be the most efficient support gas to provide a sooting plasma in these experiments [8]. It may be due to the proximity with carbon in mass and also in its lowest excitation potential of 16.6 eV being higher than the ionization potentials of carbon and its clusters. The sooting plasma so produced demonstrates a temporal growth in the densities of sputtered carbon atoms and ions as a function of the discharge voltage $V_d$ and current $I_d$. Once $C^+$ ions anchor onto a set of field contours, the direction of their consequent gyrational motion and clustering probability is determined by collision with electrons, neutral and excited C and Ne atoms. The hexapole field confinement is designed so that the radial ($B_r$), azimuthal ($B_\theta$) and axial ($B_z$) field lines produce the combined 3D magnetic field contours $-B_z(r, \theta)$ extending over the entire HC region. The streams of gyrating $C^+$ ions with large collision cross-sections eventually lead to the inside walls of HC where they impact with $E_{C^+} \approx qV_d$. The impact continuously modifies the cathode walls material characteristics and covers it with gradually increasing layers of the soot. The soot subsequently becomes the surface for later sputtering to take place. Velocity spectra of the emitted charged species have shown the dependence of carbon cluster emission on the state of sooting in the source.

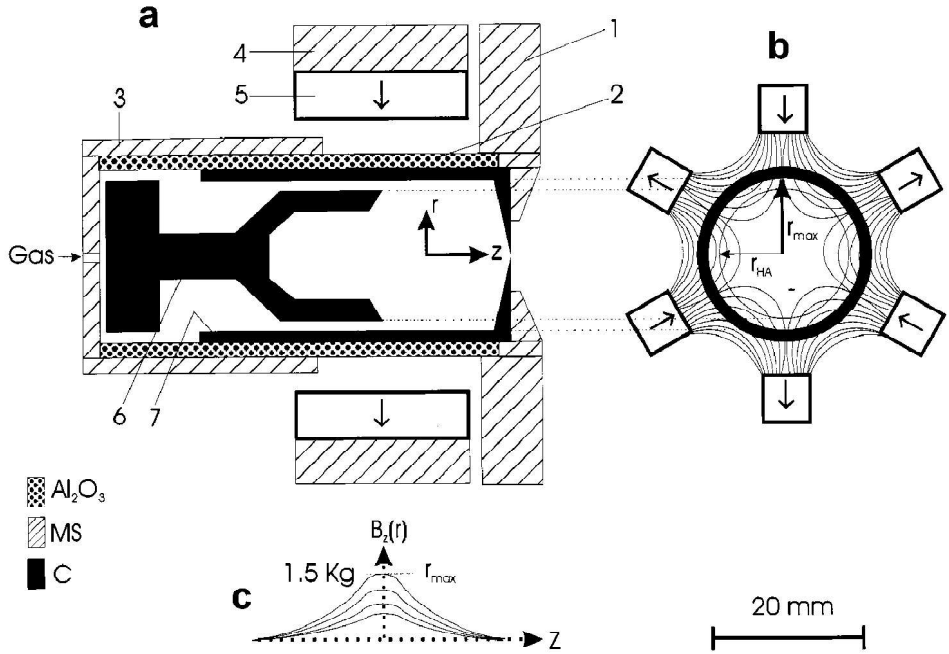

Fig. 1. The design of hollow cathode (HC) and a partially penetrating hollow anode (HA) is shown in (a) with a set of six permanent bar magnets. A mild steel (MS) base plate (#1) houses the alumina tube (#2) bonded at both ends with MS rings that hold the respective electrodes HC (#6) and HA (#7). The bar magnet set (#5) are held on the inner surface of an MS ring (#4) that acts as a field return core, magnet support and heat sink. Figure 1b shows the cusp field lines intersected by HC shown as a thick black circle at $r_{HC} = r_{max}$ and a dotted circle indicates HA's outer radius $= r_{HA}$. The 2D axial variation of $B_z(r)$ vs. r is presented in (c) where the flux density can be seen to be maximum at the centre of bar magnets. Cylindrical MS ring (#3) shapes the desired 3-dimensional $B_z(r, \theta)$ vs. z cusp field contours.

# 2 Experimental set up

The design features and operational characteristics are shown in Figure 1. The graphite hollow cathode −HC and a partially penetrating hollow anode −HA are shown in Figure 1a with the relative position of the set of six magnets. A mild steel −MS base plate holds the source which is composed of an alumina tube bonded at both ends with MS rings that hold the respective electrodes HC and HA. The hollow anode penetrates the cathodic cylinder up to the radial plane where $B_Z(r)$ has a maximum as a function of z as shown in Figure 1c. The bar magnet set are held on the inner surface of an MS ring that acts as a field return core, magnet support and a heat sink. Figure 1b shows the cross-section of the cusp field lines that are intersected by the hollow cathode at $r_{max} = r_{HC}$ and shown as black circle while a dotted circle indicates the outer radius $r_{HA}$ of the hollow anode. The axial variation of a given contour $B_Z(r)$ for different values of r as a function of z is presented in Figure 1c where the flux density can be seen to be maximum at the centre of bar magnets and the hollow anode extends to this plane. A cylindrical MS ring screwed over HA shapes the 3-dimensional $B_Z(r, \theta)$ vs. z cusp field configurations, which are seen to be crucial to the source's operational as well as clustering characteristics. Mass analysis of the extracted plasma species is performed with a permanent magnet based velocity filter. Velocity analysis has certain advantages over competitive mass analyzing techniques. Ideally a velocity spectrum contains all velocities from $v_{min}$ (≡ 0) to $v_{max}$ corresponding to masses $m_{max}$ (≡ ∞) to $m_{min}$. In practice, higher extraction voltages help to spread out the heavier masses and cluster identification improves. The analyzer's magnetic field is within easily attainable limits ~ 0.5 T on the axis.

# 3 Results

Ne+ and C+ dominate the initial velocity spectra from a newly assembled hollow cathode source. The emergence of low velocity i.e. higher mass peaks start to appear after continuous source operation after higher power ($V_d I_d$) inputs. Figure 2 was obtained after 3 hours of operation with $V_d$ = 0:85 kV, $I_d$ = 100 mA. The spectrum exhibits clusters composed of linear chains, rings and fullerenes. The higher velocity section of the spectra often consists of multiply charged species that have been partially neutralized in the extraction region, one such cluster shown is $C_3^{2+}$, otherwise, the velocity filter is not charge selective. Three groups of clusters $C_m$ can be distinguished in the spectrum; the first one is of the heavy fullerenes with m ≈ 250 to m ≥ 100, then follows the more familiar cluster range of $C_{76}$ to $C_{20}$ with main peaks around $C_{50}$ and $C_{30}$, the third regime of clusters with m < 20 to $C_1$ are dispersed over the rest of the velocity spectrum. There are two distinct stages of source operation; one is that of an un-sooted or mildly sooting stage i.e. the initial soot formative period and the other is of a well-sooted hollow cathode. Carbon cluster formation mechanisms can be directly related to the constituents of the plasma, the state of wall coverage and the discharge conditions. $V_d$ ~ 1 kV is needed for initiating the discharge at pressures ~ $10^{-1}$ mbar. A 30 min operation leads to a lower and manageable gas loads since pressure ~ $10^{-5}$ mbar is required to be maintained in the extraction chamber outside the source to avoid neutralization of the extracted clusters.

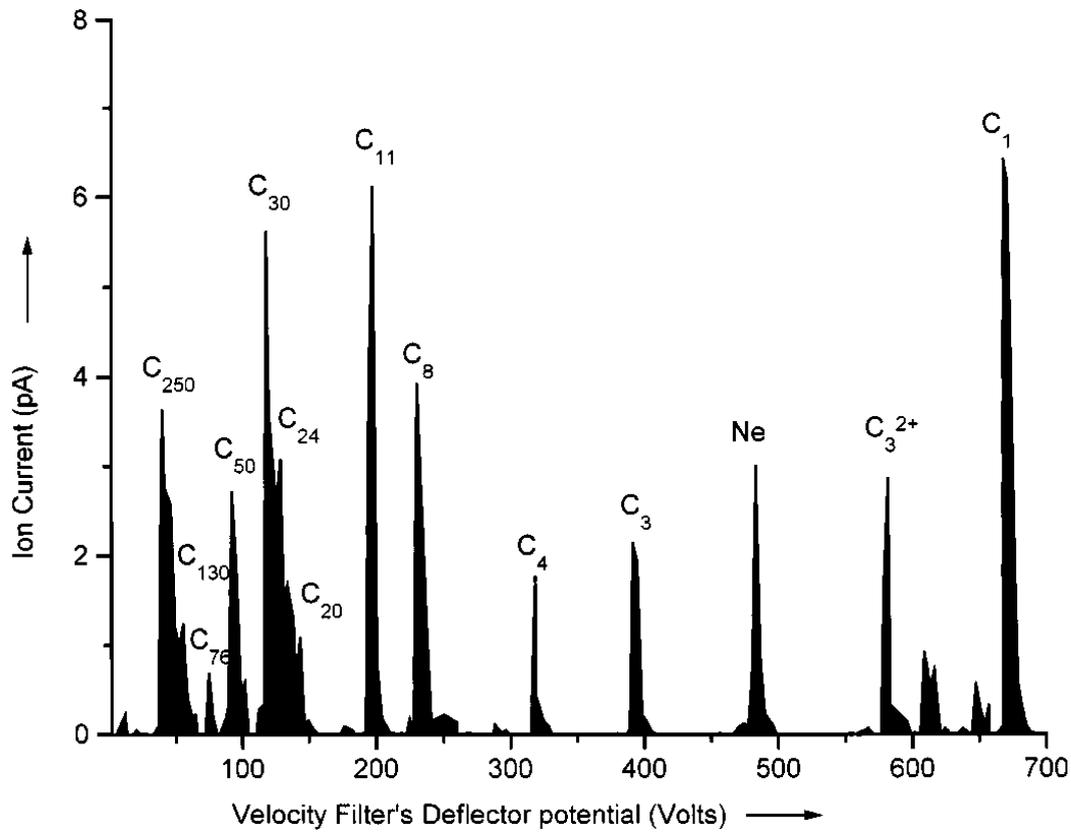

Fig. 2. This figure was obtained after 3 hours of operation with $V_d$ = 0:85 kV, $I_d$ = 100 mA and $V_{extraction}$ = 1 kV. The spectrum is composed of linear chains, rings and fullerenes. The multiply charged species in the higher velocity section of the spectra can also be seen, one such cluster shown is $C_3^{2+}$.

Figure 3a presents the velocity spectrum of the initial sooting stages of operation with Ne pressure in the source $P_g = 2-3 \times 10^{-3}$ mbar, $V_d = 0.5$ kV, $I_d = 50$ mA and $V_{extraction} = 2$ kV. Clusters ranging from $C_1$ to $C_{19}$ are present in the spectrum. On the other hand, a well-sooted source that has been operated with Ne at $V_d I_d \approx 60$ W for 20 hours yields a very different sort of spectrum shown in Figure 3b. It has been obtained with Xe as a source gas. Xe having a lower ionization potential, a broad range of excitation energies and a heavy mass was selected not to soot any further, but to utilize the existing soot on HC for the plasma's sputtered constituents. Earlier we had observed that unsooted source operating on Xe is unstable but a different behavior is exhibited by a sooted HC. The charged clusters emitted from this source have a range of fullerenes from m $\approx$ 200 to $C_{36}$ as well as the rings, chains and linear regimes of clusters. Cluster intensities are an order of magnitude higher in Figure 3b compared with those in Figure 3a.

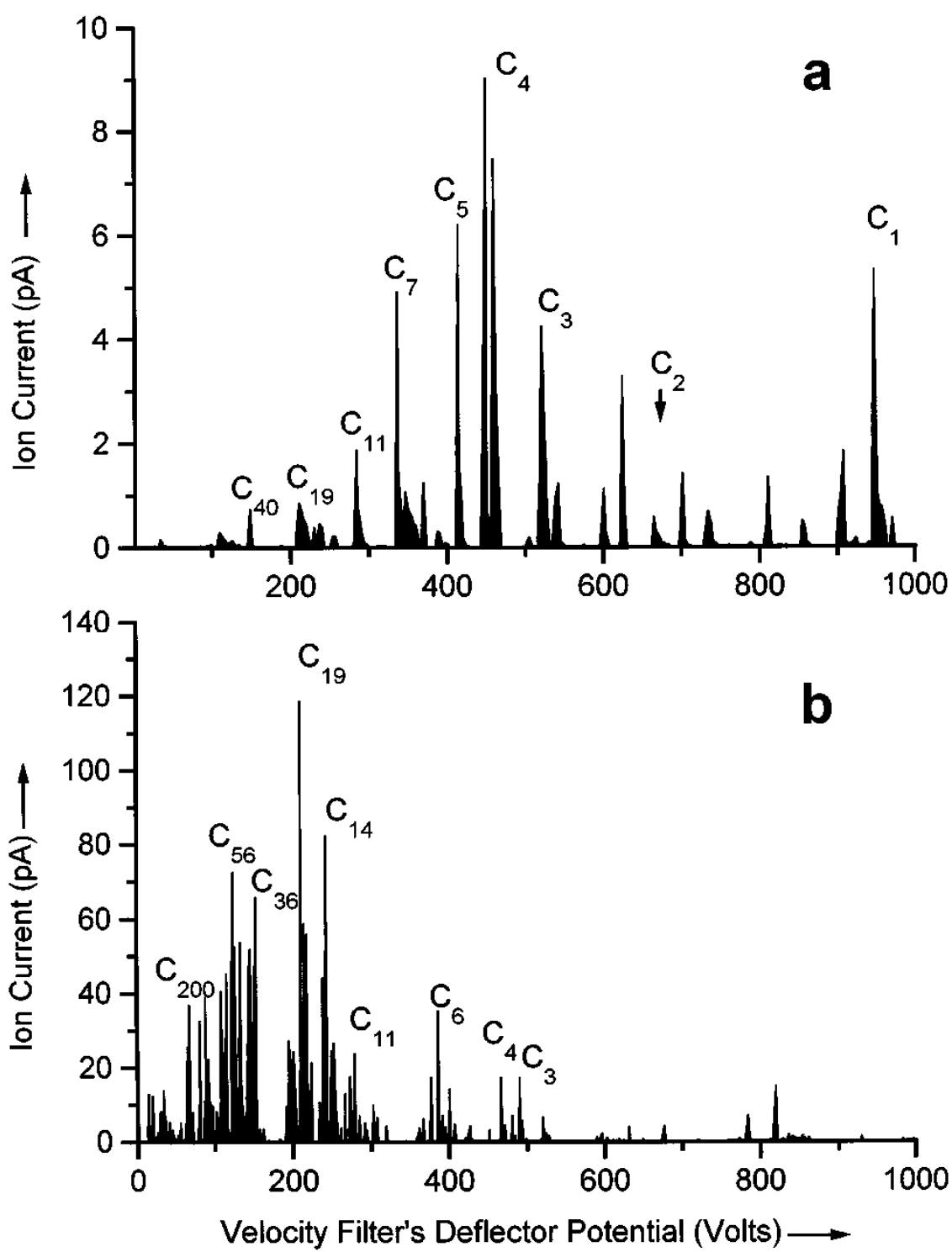

Fig. 3. The velocity spectrum of the initial sooting stages of operation with Ne is shown in (a) with pressure in the source $P_g = 2-3 \times 10^{-3}$ mbar, $V_d = 0:5$ kV, $I_d = 50$ mA and $V_{extraction} = 2$ kV. Clusters from $C_1$ to $C_{19}$ are present in the spectrum. Figure (b) shows results from a well-sooted source operated with Ne at ≈ 60 W for 20 hours. It has been obtained with Xe as a source gas. The emitted charged clusters have a range of fullerenes from m ≈ 200 to $C_{36}$ as well as the rings, chains and linear regimes of clusters. Note the difference in cluster ion intensities in the two spectra.

# 4 Conclusions

Our study with different structural designs has shown that the hollow cathode source with cusp magnetic field has two distinct discharge regions for all configurations; the first is the narrow annular region between the outer walls of HA and the inside of HC extending from the tip of HA backwards. This is an annular disc of 2 mm thickness and 15—20 mm length within $B_Z (r, \theta) = 1.0$—$1.5$ kgauss. The $B_Z (r, \theta)$ cusps and edges produce 3 major elliptic zones on the outside of HA due to the impact of electrons and three regions on the inside of HC. The HC elliptic zones have enhanced sputtering activity due to the field directed $Ne^+$ and $C^+(m \geq 1)$ impact. The cathodic elliptic areas are also the regions of electron emission, high collision activity, the removal of wall material and possibly ion-induced clustering in the sooted layers. Another cylindrical region of much larger dimensions compared with the one discussed earlier is outside the HA to the HC's cylindrical region with the $r_{HC} = r_{max}$ and length $= 20 \pm 5$ mm. It has the complete range of cusps and edges of the $B_Z (r, \theta)$ field. The larger variation in $B_Z (r, \theta)$ along with a plasma filling the entire region indicates inherent clustering environment where a broad range of collisions between the plasma species are possible. An interesting result from our cusp field, glow discharge plasma is the relatively reduced peak intensities of $C_{60}$. Search for the elusive $C_{60}$ as a function of various source parameters has forced us to draw the conclusion that probably, due to the inherent spin, its magnetic moment strongly interacts with $B_Z (r, \theta)$. It may have been produced by the same cluster mechanisms as are prevalent for other clusters but remains anchored to the respective field contours not only in the charged states but also in the neutral modes. Alternatively, it can be argued that the magnetic field may be a barrier to the very formation of the Buckyball.